\documentclass[12pt]{article}
\usepackage{amssymb}
\usepackage{color}
\usepackage{graphicx}

\newcommand{\A}{{\mathfrak A}}

\newcommand{\mc}{\mathcal}

\newcommand{\be}{\begin{equation}}
\newcommand{\en}{\end{equation}}
\newcommand{\bea}{\begin{eqnarray}}
\newcommand{\ena}{\end{eqnarray}}

\newcommand{\FF}{\mc F}
\newcommand{\GG}{\mc G}
\newcommand{\LL}{\mc L}

\newcommand{\1}{1 \!\! 1}
\newcommand{\ST}{\mc S}

\newcommand{\Hil}{\mc H}

\catcode `\@=11 \@addtoreset{equation}{section}

\catcode `\@=12

\begin{document}

\begin{center}
{\Large \bf A quantum statistical approach to simplified stock markets}   \vspace{2cm}\\

{\large F. Bagarello}
\vspace{3mm}\\
  Dipartimento di Metodi e Modelli Matematici,
Facolt\`a di Ingegneria,\\ Universit\`a di Palermo, I - 90128  Palermo, Italy\\
E-mail: bagarell@unipa.it\\home page:
www.unipa.it$\backslash$\~\,bagarell
\vspace{2mm}\\
\end{center}

\vspace*{2cm}

\begin{abstract}
\noindent We use standard perturbation techniques originally formulated in quantum (statistical) mechanics in the analysis of a toy model of a stock market which is given in terms of bosonic operators. In particular we discuss the probability of transition from a given value of the {\em portfolio} of a certain trader to a different one. This computation  can also be carried out using some kind of {\em Feynman graphs} adapted to the present context.

\end{abstract}

\vfill

\newpage

\section{Introduction and motivations}

In some recent papers, \cite{bag1,bag2,bag3}, we have discussed why and
how a quantum mechanical framework, and in particular operator algebras and the number representation, can be used in the analysis of
 some simplified models of stock  markets. These models are just prototypes of real stock markets because, among the other simplifications, we are not considering financial derivatives. For this reason our interest looks different from that widely discussed in \cite{baa}, even if the general settings appear to be very close (and very close also to the framework used in \cite{scha1}).
The main reason for using operator algebras in the analysis of these simplified
closed stock markets comes from the following considerations: in the
closed market we have in mind the total amount of cash stays constant. Also, the
total number of shares does not change with time. Moreover, when a
trader $\tau$ interacts with a second trader $\sigma$, they change
money and shares in a {\em discrete fashion}: for instance,  $\tau$ increments his number of shares of 1 unit while his cash
decrements of a certain number of {\em monetary units} (which is the
minimum amount of cash existing in the market: 1 cent of dollar, for
example), which is exactly the price of the share. Of course, for the trader $\sigma$ the situation is just
reversed. So we have at least two quantities, the cash and the
number of shares, which change discontinuously as multiples of two fixed
quantities. In \cite{bag1,bag2,bag3} we also have two other quantities defining our simplified
market: the price of the share (in the cited papers the traders
can exchange just a single kind of shares!) and the {\em market
supply}, i.e. the overall tendency of the market to sell a share.
It is clear that also the price
of the share must change discontinuously, and that's why we have assumed that also the market supply is labeled by a discrete quantity.

Operator algebras and quantum statistical mechanics provide a very natural
settings for discussing such a system. Indeed they produce a natural
way for: (a) describing quantities which change with discrete steps;
(b) obtaining the differential equations for the relevant variables
of the system under consideration, the so-called {\em observables of
the system}; (c) finding conserved quantities; (d) compute transition probabilities.

For these reasons we have suggested in \cite{bag1,bag2,bag3}  an operator-valued scheme for the description of such a simplified market.
Let us see why, neglecting here all the many mathematical
complications arising mainly from the fact that our operators are
unbounded, and limiting our introduction to few important facts in
quantum mechanics and second quantization which will be used in the next sections. More details can be
found, for instance, in \cite{mer,reed} and \cite{brat}, as well as in \cite{bag1,bag2,bag3}.

Let $\Hil$ be an Hilbert space and $B(\Hil)$ the set of all the
bounded operators on $\Hil$.   Let $\ST$ be
our (closed) physical system and $\A$ the set of all the operators, which may be unbounded, useful for
a complete description of $\ST$, which includes the { observables } of $\ST$.  The description of the time evolution of
$\ST$ is driven by a self-adjoint operator $H=H^\dagger$ which is
called {\em the hamiltonian} of $\ST$ and which in standard quantum
mechanics represents  the energy of $\ST$. In the {\em Heisenberg} picture the time
evolution of an observable $X\in\A$ is given by \be
X(t)=e^{iHt}Xe^{-iHt}\label{a1}\en or, equivalently, by the solution
of the differential equation \be
\frac{dX(t)}{dt}=ie^{iHt}[H,X]e^{-iHt}=i[H,X(t)],\label{a2}\en where
$[A,B]:=AB-BA$ is the {\em commutator } between $A$ and $B$. The
time evolution defined in this way is usually a one parameter group
of automorphisms of $\A$. The wave function $\Psi$ of $\ST$ is constant in time.

In the Scr\"odinger picture the situation is just reversed: an observable $X\in\A$ does not evolve in time (but if it has some explicit dependence on $t$) while the wave function $\Psi$ of $\ST$ satisfies the Scr\"odinger equation $
i\,\frac{\partial \Psi(t)}{\partial t}=H\,\Psi(t)$, whose formal solution is, if $H$ does not depend on time, $\Psi(t)=e^{-iHt}\Psi$.

In our paper a special role is played by the so called {\em
canonical commutation relations } (CCR): we say that a set of
operators $\{a_l,\,a_l^\dagger, l=1,2,\ldots,L\}$ satisfy the CCR if
the following hold:\be [a_l,a_n^\dagger]=\delta_{ln}\1,\hspace{8mm}
[a_l,a_n]=[a_l^\dagger,a_n^\dagger]=0 \label{a3}\en for all
$l,n=1,2,\ldots,L$. Here $\1$ is the identity operator of $B(\Hil)$. These
operators, which are widely analyzed in any textbook in quantum
mechanics, see \cite{mer} for instance, are those which are used to
describe $L$ different {\em modes} of bosons. From these operators
we can construct $\hat n_l=a_l^\dagger a_l$ and $\hat N=\sum_{l=1}^L
\hat n_l$ which are both self-adjoint. In particular $\hat n_l$ is
the {\em number operator } for the l-th mode, while $\hat N$ is the
{\em number operator of $\ST$}.

The Hilbert space of our system is constructed as follows: we
introduce the {\em vacuum} of the theory, that is a vector
$\varphi_0$ which is annihilated by all the operators $a_l$:
$a_l\varphi_0=0$ for all $l=1,2,\ldots,L$. Then we act on
$\varphi_0$ with the  operators $a_l^\dagger$ and their powers: \be
\varphi_{n_1,n_2,\ldots,n_L}:=\frac{1}{\sqrt{n_1!\,n_2!\ldots
n_L!}}(a_1^\dagger)^{n_1}(a_2^\dagger)^{n_2}\cdots
(a_L^\dagger)^{n_L}\varphi_0, \label{a4}\en $n_l=0,1,2,\ldots$ for all $l$. These vectors form an
orthonormal set and are eigenstates of both $\hat n_l$ and $\hat N$:
$\hat
n_l\varphi_{n_1,n_2,\ldots,n_L}=n_l\varphi_{n_1,n_2,\ldots,n_L}$ and
$\hat N\varphi_{n_1,n_2,\ldots,n_L}=N\varphi_{n_1,n_2,\ldots,n_L}$,
where $N=\sum_{l=1}^Ln_l$. Moreover using the  CCR we deduce that
$\hat
n_l\left(a_l\varphi_{n_1,n_2,\ldots,n_L}\right)=(n_l-1)(a_l\varphi_{n_1,n_2,\ldots,n_L})$
and $\hat
n_l\left(a_l^\dagger\varphi_{n_1,n_2,\ldots,n_L}\right)=(n_l+1)(a_l^\dagger\varphi_{n_1,n_2,\ldots,n_L})$,
for all $l$. For these reasons the following interpretation is
given: if the $L$ different modes of bosons of $\ST$ are described
by the vector $\varphi_{n_1,n_2,\ldots,n_L}$, this implies that
$n_1$ bosons are in the first mode, $n_2$ in the second mode, and so
on. The operator $\hat n_l$ acts on $\varphi_{n_1,n_2,\ldots,n_L}$
and returns $n_l$, which is exactly the number of bosons in the l-th
mode. The operator $\hat N$ counts the total number of bosons.
Moreover, the operator $a_l$ destroys a boson in the l-th mode,
while $a_l^\dagger$ creates a boson in the same mode. This is why $a_l$ and
$a_l^\dagger$ are usually called the {\em annihilation} and the {\em
creation} operators.

The Hilbert space $\Hil$ is obtained by taking the closure of the linear
span of all the vectors in (\ref{a4}).

 \vspace{2mm}

An operator $Z\in\A$ is a {\em constant of motion} if it commutes
with $H$. Indeed in this case equation (\ref{a2}) implies that $\dot
Z(t)=0$, so that $Z(t)=Z$ for all $t$.

The vector $\varphi_{n_1,n_2,\ldots,n_L}$ in (\ref{a4}) defines a
{\em vector (or number) state } over the algebra $\A$  as
\be\omega_{n_1,n_2,\ldots,n_L}(X)=
<\varphi_{n_1,n_2,\ldots,n_L},X\varphi_{n_1,n_2,\ldots,n_L}>,\label{a5}\en
where $<\,,\,>$ is the scalar product in $\Hil$. As we have discussed in \cite{bag1,bag2}, these states
may be used to {\em project} from quantum to classical dynamics and to
fix the initial conditions of the market.

\vspace{2mm}

The paper is organized as follows:

In  Section II we introduce a new model, slightly different from the one proposed in \cite{bag2}, and we deduce some of its features and the related equations of motion, working in the Heisenberg picture. One of the main improvements with respect to \cite{bag2} is that several kind of shares (and not just one!) will be considered here.

In Section III we adopt a different point of view, using the Scr\"odinger picture to deduce the transition probability from a given initial situation to a final state, corresponding to two different values of the portfolios of the various traders. Since the reader might not be familiar with the tools adopted, we will be rather explicit in the derivation of our results.

Section IV is devoted to  the
conclusions and to our plans for the future.

\section{The  model and first considerations}

Let us consider $N$ different traders $\tau_1$, $\tau_2$, $\ldots$, $\tau_N$, exchanging $L$ different kind of shares $\sigma_1$, $\sigma_2$, $\ldots$, $\sigma_L$. Each trader has a starting amount of cash, which is used during the trading procedure: the cash of the trader who sells a share increases while the cash of the trader who buys that share consequently decreases. The absolute value of these variations is the price of the share at the time in which the transaction takes place. Following our previous results we start introducing a set of bosonic operators which are listed, together with their economical meaning, in the following table. We  adopt here latin indexes to label the traders and greek indexes for the shares: $j=1,2,\ldots,N$ and $\alpha=1,2,\ldots,L$.

\vspace{5mm}
{\hspace{-.8cm}
\begin{tabular}{|c||c||c||c||c|} \hline    &the operator and.. &...its economical meaning               \\
\hline  & $a_{j,\alpha}$  & annihilates a share $\sigma_\alpha$  in the portfolio of $\tau_j$    \\
\hline   & $a_{j,\alpha}^\dagger$ & creates a share $\sigma_\alpha$  in the portfolio of $\tau_j$   \\
\hline  & $\hat n_{j,\alpha}=a_{j,\alpha}^\dagger a_{j,\alpha}$ &counts the number of share $\sigma_\alpha$  in the portfolio of $\tau_j$   \\
\hline\hline
\hline  & $c_j$  & annihilates a monetary unit in the portfolio of $\tau_j$    \\
\hline   & $c_j^\dagger$ & creates a monetary unit in the portfolio of $\tau_j$   \\
\hline  & $\hat k_j=c_j^\dagger c_j$ &counts the number of monetary units in the portfolio of $\tau_j$   \\
\hline\hline
\hline  & $p_\alpha$  &  lowers the price of the share $\sigma_\alpha$ of one unit of cash   \\
\hline   & $p_\alpha^\dagger$ &  increases the price of the share $\sigma_\alpha$ of one unit of cash   \\
\hline  & $\hat P_\alpha=p_\alpha^\dagger p_\alpha$ &gives  the  value of the share $\sigma_\alpha$  \\
\hline\hline

\end{tabular}

\vspace{4mm}

\indent
Table 1.-- List of operators and of their {\em economical} meaning.}

\vspace{3mm}

These operators are bosonic in the sense that they satisfy the following commutation rules
\be [c_j,c_k^\dagger]=\1\,\delta_{j,k},\quad [p_\alpha,p_\beta^\dagger]=\1\,\delta_{\alpha,\beta}\quad [a_{j,\alpha},a_{k,\beta}^\dagger]=\1\,\delta_{j,k}\delta_{\alpha,\beta},\label{21}\en

while all the other commutators are zero.

As discussed in the Introduction, in \cite{bag2,bag3} we have also introduced another set of operators related to the {\em market supply} which was used to deduce the dynamics of the price of the single kind of share considered there. However, the mechanism proposed in those papers, thought being reasonable, is too naive and gives no insight on the nature of the market itself. For this reason in \cite{bag3,bagdmmm} we have also considered a different point of view, leaving open the problem of finding the dynamics of the price and focusing the attention on the time evolution of the portfolio of a fixed trader. This is the same point of view which we briefly consider in this section, while we will comment on other possibilities in the rest of the paper. More precisely, as we have discussed in the Introduction, the dynamical behavior of our market is driven by a certain hamiltonian $\hat H$. We assume that $\hat H$ can be written as $\hat H=H+H_{prices}$, where
 \be
\left\{
\begin{array}{ll}
H=H_0+ \lambda\,H_I, \mbox{ with }  \\
H_0 = \sum_{j,\alpha}\,\omega_{j,\alpha}\, \hat n_{j,\alpha}+\sum_{j}\,\omega_j\, \hat k_j\\
 H_I = \sum_{i,j,\alpha}\,p_{i,j}^{(\alpha)}\left(a_{i,\alpha}^\dagger a_{j,\alpha}c_i^{\hat P_\alpha} {c_j^\dagger}^{\hat P_\alpha}+h.c.\right). \\
\end{array}
\right. \label{22} \en Here h.c. stands for hermitian conjugate, $c_i^{\hat P_\alpha}$ and ${c_j^\dagger}^{\hat P_\alpha}$ are defined as in \cite{bag2}, and $\omega_{j,\alpha}$,  $\omega_j$ and $p_{i,j}^{(\alpha)}$
are positive real numbers. In particular these last coefficients assume different values depending on the possibility of $\tau_i$ to interact with $\tau_j$ and exchanging a share $\sigma_\alpha$: for instance $p_{2,5}^{(1)}=0$ if there is no way for $\tau_2$ and $\tau_5$ to exchange a share $\sigma_1$. Notice that this does not exclude that, for instance, they could exchange a share $\sigma_2$, so that $p_{2,5}^{(2)}\neq 0$.  With this in mind it is natural to put $p_{i,i}^{(\alpha)}=0$ and $p_{i,j}^{(\alpha)}=p_{j,i}^{(\alpha)}$.

Going back to (\ref{22}), we observe that $H_0$ is nothing but the standard free hamiltonian which is used for many-body systems like the ones we are considering here (where the {\em bodies} are nothing but the {\em traders}, and the {\em cash}). More interesting is the meaning of the interaction hamiltonian $H_I$. To understand $H_I$ we consider its action on a vector like
\be
\varphi_{\{n_{j,\alpha}\};\{k_j\};\{P_\alpha\}}:=\frac{{a_{1,1}^\dagger}^{\!\!\!n_{1,1}}\,\cdots {a_{N,L}^\dagger}^{\!\!\!n_{N,L}} {c_1^\dagger}^{k_1}\,\cdots {c_N^\dagger}^{k_N}
{p_1^\dagger}^{P_1}\cdots{p_L^\dagger}^{P_L}}{\sqrt{n_{11}!\cdots n_{N,L}!k_1!\cdots k_L!P_1!\cdots P_L!}}\,\varphi_0, \label{23}\en
where  $\varphi_0$ is the vacuum of all the annihilation operators involved here, see Section I.

Because of the CCR we deduce that the action of a single contribution of $H_I$, $a_{i,\alpha}^\dagger a_{j,\alpha}c_i^{\hat P_\alpha} {c_j^\dagger}^{\hat P_\alpha}$, on $\varphi_{\{n_{j,\alpha}\};\{k_j\};\{P_\alpha\}}$ is proportional to another vector $\varphi_{\{n_{j,\alpha}'\};\{k_j'\};\{P_\alpha'\}}$ with just 4 different quantum numbers. In particular $n_{j,\alpha}$, $n_{i,\alpha}$, $k_j$ and $k_i$ are replaced respectively by $n_{j,\alpha}-1$, $n_{i,\alpha}+1$, $k_j+P_\alpha$ and $k_i-P_\alpha$ (if this is larger or equal than zero, otherwise the vector is annihilated). This means that $\tau_j$ is selling a share $\sigma_\alpha$ to $\tau_i$ and earning money from this operation. For this reason it is convenient to introduce the following {\em selling} and {\em buying} operators:
\be
x_{j,\alpha}:=a_{j,\alpha}\,{c_j^\dagger}^{\hat P_\alpha},\qquad x_{j,\alpha}^\dagger:=a_{j,\alpha}^\dagger\,{c_j}^{\hat P_\alpha}
\label{24}\en
With these definitions and using the properties of the coefficients $p_{i,j}^{(\alpha)}$ we can rewrite $H_I$ as
\be
 H_I = 2\,\sum_{i,j,\alpha}\,p_{i,j}^{(\alpha)}x_{i,\alpha}^\dagger\,x_{j,\alpha} \Rightarrow H=\sum_{j,\alpha}\,\omega_{j,\alpha}\, \hat n_{j,\alpha}+\sum_{j}\,\omega_j\, \hat k_j+2\,\lambda\,\sum_{i,j,\alpha}\,p_{i,j}^{(\alpha)}x_{i,\alpha}^\dagger\,x_{j,\alpha}
\label{25}\en
In \cite{bagdmmm} we have discussed the role of  $H_{price}$ which should be used to deduce the time evolution of the operators $\hat P_\alpha$, $\alpha=1,2,\ldots,L$, and which will not be fixed in these notes. This will be justified below, after getting the differential equations of motion for our system. Again in \cite{bagdmmm} we have also shown that $\hat H$ corresponds to a closed market where the money and the total number of shares of each type are conserved. Indeed, calling $\hat N_\alpha:=\sum_{l=1}^N\hat n_{l,\alpha}$ and $\hat K:=\sum_{l=1}^N\hat k_{l}$ we have seen that, for all $\alpha$,
$[\hat H,\hat N_\alpha]=[\hat H,\hat K]=0$. Hence $\hat N_\alpha$ and $\hat K$  are integrals of motion, as expected. Of course, something different may happen in realistic markets. For instance, a given company could decide to issue more stocks in the market or to split existing stocks. The related $\hat N_\alpha$, say $\hat N_{\alpha_0}$, is no longer a constant of motion. Hence $\hat H$ should be modified in order to have $[\hat H,\hat N_{\alpha_0}]\neq 0$. For that it is enough to add in $\hat H$ some {\em source} or {\em sink} contribution, trying to preserve the self-adjointness of $\hat H$. However, this is not always the more natural choice. For instance, if we want to include in the model interactions with the environment (economical, political, social inputs), it could be more convenient to use non-hermitean operators, like the generators appearing in the analysis of quantum dynamical semigroups used to describe open systems, \cite{alicki}. This aspect will not be considered here.

Let us now define the {\em portfolio operator} of the trader $\tau_l$ as
\be \hat
\Pi_l(t)=\sum_{\alpha=1}^L\hat P_\alpha(t)\,\hat n_{l,\alpha}(t)+\hat k_l(t).\label{27}\en
This is a natural definition, since it is just the sum of the cash and of the total value of the shares that $\tau_l$ possesses at time $t$. Once again, we stress that in our simplified model there is no room for the financial derivatives.

Using (\ref{a2}) and the commutation rules assumed so far we derive the following system of equations:
\be\left\{\begin{array}{lll}\frac{d\hat
n_{l,\alpha}(t)}{dt}=2i\lambda\,\sum_{j=1}^N\,p_{i,j}^{(\alpha)}\,\left(x_{j,\alpha}^\dagger(t)\,x_{l,\alpha}(t)-
x_{l,\alpha}^\dagger(t)\,x_{j,\alpha}(t)\right),\\
\frac{d\hat k_l(t)}{dt}=-2i\lambda\,\sum_{j=1}^N\,\sum_{\alpha}^L\,p_{i,j}^{(\alpha)}\,\hat P_\alpha(t)\,\left(x_{j,\alpha}^\dagger(t)\,x_{l,\alpha}(t)-
x_{l,\alpha}^\dagger(t)\,x_{j,\alpha}(t)\right),\\
\frac{dx_{l,\alpha}(t)}{dt}=i\,x_{l,\alpha}(t)\,(\omega_{l}\, \hat P_\alpha(t)-\omega_{l,\alpha})+\\
\hspace{15mm}+2i\lambda\sum_{j=1}^N\sum_{\beta=1}^L\,p_{l,j}^{(\beta)}\,
[x_{l,\beta}(t)^\dagger(t),x_{l,\alpha}(t)]\,x_{j,\beta}(t),\\
\end{array}\right.\label{28}\en
which, together with their adjoints, produce a closed system of differential equation. Notice that these equations imply that $\sum_{\alpha=1}^L \hat P_\alpha(t)\, \frac{d}{dt}\hat n_{l,\alpha}(t)+ \frac{d}{dt}\hat k_l(t)=0$. This system is now replaced by a semi-classical approximation which is obtained replacing the time dependent operators $\hat P_\alpha(t)$ with $L$ classical fields $P_\alpha(t)$ which are deduced by empirical data. This is the reason why we were not interested in fixing $H_{prices}$ in $\hat H$: $\hat P_\alpha(t)$ is replaced by $P_\alpha(t)$ which are no-longer internal degrees of freedom of the model but, rather than this, simple external classical fields.
Neglecting all the details, which can be found in \cite{bagdmmm}, we observe that the first non trivial contribution in $\lambda$ is
\be\left\{\begin{array}{lll}
n_{l,\alpha}(t):=\omega_{\{n_{j,\alpha}\};\{k_j\};\{P_\alpha\}}(\hat n_{l,\alpha}(t))=\\
=n_{l,\alpha}-
8\lambda^2\,\sum_{j=1}^N\,\left(p_{l,j}^{(\alpha)}\right)^2\,\tilde M_{j,l;\alpha}\,\Re(\Theta^{(2)}_{j,l;\alpha}(t))=:n_{l,\alpha}+\delta n_{l,\alpha}(t),\\
k_{l}(t):=\omega_{\{n_{j,\alpha}\};\{k_j\};\{P_\alpha\}}(\hat k_{l}(t))=\\
=k_{l}+
8\lambda^2\,\sum_{j=1}^N\,\sum_{\alpha=1}^L\,\left(p_{l,j}^{(\alpha)}\right)^2\,\tilde M_{j,l;\alpha}\,\Re(\Theta^{(3)}_{j,l;\alpha}(t))=:k_{l}+\delta k_{l}(t),\\
\end{array}\right.\label{29}\en
where we have defined:
\be\left\{\begin{array}{lll}
\tilde M_{j,l;\alpha}=M_{j,l;\alpha}-M_{l,j;\alpha},\\
M_{j,l;\alpha}=n_{j,\alpha}n_{l,\alpha}\frac{(k_j+P_\alpha)!}{k_j!}\,\frac{(k_l+P_\alpha)!}{k_l!}-
n_{j,\alpha}(1+n_{l,\alpha})\frac{(k_j+P_\alpha)!}{k_j!}\,\frac{k_l!}{(k_l-P_\alpha)!},\\
\end{array}\right.\label{210}\en
and
\be\left\{\begin{array}{lll}
\Theta^{(3)}_{j,l;\alpha}(t)=\int_0^t\,P_\alpha(t')\,\Theta^{(1)}_{j,l;\alpha}(t')e^{-i\,\Theta^{(0)}_{j,l;\alpha}(t')}
\,dt',\\
\Theta^{(2)}_{j,l;\alpha}(t):=\int_0^t\,\Theta^{(1)}_{j,l;\alpha}(t')e^{-i\,\Theta^{(0)}_{j,l;\alpha}(t')}\,dt',\\
\Theta^{(1)}_{j,l;\alpha}(t):=\int_0^t\,e^{-i\,\Theta^{(0)}_{j,l;\alpha}(t')}\,dt',\\
\Theta^{(0)}_{j,l;\alpha}(t):=(\omega_j-\omega_l)\,\int_0^t\,P_\alpha(t')\,dt'-(\omega_{j,\alpha}-\omega_{l,\alpha})t\\
\end{array}\right.\label{211}\en

The time dependence of the portfolio can now be written as \be
\Pi_l(t):=\omega_{\{n_{j,\alpha}\};\{k_j\};\{P_\alpha\}}(\hat \Pi_l(t))=\Pi_l(0)+\delta\Pi_l(t),\label{212}\en with  \be
\delta\Pi_l(t)=\sum_{\alpha=1}^L\,n_{l,\alpha}(P_\alpha(t)-P_\alpha(0))+ \sum_{\alpha=1}^L\,P_\alpha(t)\,\delta n_{l,\alpha}(t)+\delta k_l(t).\label{213}\en
It should be emphasized that, even under the approximations we are considering here, we still find $\sum_{\alpha=1}^L P_\alpha(t)\,\dot n_{l,\alpha}(t)+\dot k_l(t)=0$. However, we are loosing the conservation laws we have discussed before: both $\hat N_l$ and $\hat K$ do not commute with the effective hamiltonian which produces the semi-classical version of (\ref{28}) anymore, and therefore they are not constant in time. This imposes some strict constraint on the validity of our expansion, as we have discussed already in \cite{bag3} and suggests the different approach to the problem which we will discuss in the next section. Again we refer to \cite{bagdmmm} for more comments on these results.

\section{A time dependent point of view}

In this section we will consider a slightly different point of view. Our hamiltonian has no $H_{price}$ contribution at all, since the price operators $\hat P_\alpha$, $\alpha=1,\ldots,L$, are now replaced from the very beginning by external classical fields $ P_\alpha(t)$, whose time dependence describes, as an input of the model, the variation of the prices of the shares. Incidentally, this implies that possible fast changes of the prices are automatically included in the model through the analytic expressions of the functions $ P_\alpha(t)$.  Hence the interaction hamiltonian $H_I$  in (\ref{25}) turns out to be a time dependent operator, $H_I(t)$. More in details, the hamiltonian $H=H_0+\lambda H_I(t)$ of the model looks like the one in (\ref{25}) but with the following time-dependent selling and buying operators:
\be
x_{j,\alpha}(t):=a_{j,\alpha}\,{c_j^\dagger}^{P_\alpha(t)},\qquad x_{j,\alpha}^\dagger(t):=a_{j,\alpha}^\dagger\,{c_j}^{P_\alpha(t)}
\label{31}\en
The $L$ price functions $P_\alpha(t)$ will be taken piecewise constant, since the price of a share changes discontinuously: it has a certain value before the transaction and (in general) a different value after the transaction. This new value does not change until the next transaction takes place. More in details, we introduce a time step $h$ which we call {\em the time of transaction}, and we divide the interval $[0,t[$ in subintervals of duration $h$: $[0,t[=[t_0,t_1[\cup[t_0,t_1[\cup[t_1,t_2[\cdots[t_{M-1},t_M[$, where $t_0=0$, $t_1=h$, $\ldots$, $t_{M-1}=(M-1)h=t-h$, $t_M=Mh=t$. Hence $h=t/M$. As for the prices, for $\alpha=1,\ldots L$ we put
\be P_\alpha(t)=\left\{\begin{array}{lll}
P_{\alpha,0},\hspace{25mm} t\in [t_0,t_1[,\\
P_{\alpha,1},\hspace{25mm} t\in [t_1,t_2[,\\
\ldots\ldots,\\
P_{\alpha,M-1},\hspace{19mm} t\in [t_{M-1},t_M[\\
\end{array}\right.\label{32}\en
An orthonormal basis in the Hilbert space of the model $\Hil$ is now the set of vectors defined as
\be
\varphi_{\{n_{j,\alpha}\};\{k_j\}}:=\frac{{a_{1,1}^\dagger}^{n_{1,1}}\,\cdots {a_{N,L}^\dagger}^{n_{N,L}} {c_1^\dagger}^{k_1}\,\cdots {c_N^\dagger}^{k_N}
}{\sqrt{n_{11}!\cdots n_{N,L}!k_1!\cdots k_L!}}\,\varphi_0, \label{33}\en
where  $\varphi_0$ is the vacuum of all the annihilation operators involved here. They differ from the ones in (\ref{23}) since the price operators disappear, of course. To simplify the notation  we introduce a set $\FF=\{\{n_{j,\alpha}\};\{k_j\}\}$ so that the vectors of the basis will be simply written as $\varphi_\FF$.

The main problem we want to discuss here is the following: suppose that at $t=0$ the market is described by a vector
$\varphi_{\FF_0}$. This means that, since $\FF_0=\{\{n_{j,\alpha}^o\},\{k_j^o\}\}$, at $t=0$ the trader $\tau_1$ has $n_{11}^o$ shares of $\sigma_1$, $n_{12}^o$ shares of $\sigma_2$, $\ldots$, and $k_1^o$ units of cash. Analogously, the trader $\tau_2$ has $n_{21}^o$ shares of $\sigma_1$, $n_{22}^o$ shares of $\sigma_2$, $\ldots$, and $k_2^o$ units of cash. And so on. We want to compute the probability that at time $t$ the market has moved to the configuration $\FF_f=\{\{n_{j,\alpha}^f\},\{k_j^f\}\}$. This means that, for example, $\tau_1$ has now $n_{11}^f$ shares of $\sigma_1$, $n_{12}^f$ shares of $\sigma_2$, $\ldots$, and $k_1^f$ units of cash.

Similar problems are very well known in ordinary quantum mechanics: we need to compute a probability transition from the original state  $\varphi_{\FF_0}$ to a final state $\varphi_{\FF_f}$, and therefore we will use here the standard time-dependent perturbation scheme for which we refer to \cite{mess}. The main difference with respect to what we have done in the previous section is the use of the Schr\"odinger rather than the Heisenberg picture. Hence the market is described by a time-dependent wave function $\Psi(t)$ which, for $t=0$, reduces to $\varphi_{\FF_0}$: $\Psi(0)=\varphi_{\FF_0}$. The transition probability we are looking for is
\be
P_{\FF_0\rightarrow\FF_f}(t):=\left|<\varphi_{\FF_f},\Psi(t)>\right|^2
\label{34}\en
The computation of $P_{\FF_0\rightarrow\FF_f}(t)$ is a standard exercise, \cite{mess}. In order to make the paper accessible also to those people who are not familiar with quantum mechanics, we give here the main steps of its derivation.

Since the set of the vectors $\varphi_{\FF}$ is an orthonormal basis in $\Hil$ the wave function $\Psi(t)$ can be written as
\be
\Psi(t)=\sum_\FF c_\FF(t)\,e^{-iE_\FF t}\varphi_{\FF},
\label{35}\en
where $E_\FF$ is the eigenvalue of $H_0$ defined as
\be
H_0\varphi_{\FF}=E_{\FF}\varphi_{\FF}, \qquad\Rightarrow \qquad E_\FF=\sum_{j,\alpha}\omega_{j,\alpha}n_{j,\alpha}+\sum_{j}\omega_{j}k_{j}.
\label{36}\en
This is a consequence of the fact that $\varphi_{\FF}$ in (\ref{33}) is an eigenstate of $H_0$ in (\ref{22}). Using the quantum mechanical terminology, we sometimes call $E_\FF$ the {\em free energy} of $\varphi_\FF$.
Putting (\ref{35}) in (\ref{34}), and recalling that $<\varphi_{\FF},\varphi_{\GG}>=\delta_{\FF,\GG}$, we have
\be
P_{\FF_0\rightarrow\FF_f}(t):=\left|c_{\FF_f}(t)\right|^2
\label{37}\en
The answer to our original question is therefore given if we are able to compute $c_{\FF_f}(t)$ in (\ref{35}). Due to the analytic form of our hamiltonian, this cannot be done exactly. However, several possible perturbation schemes exist in the literature. We will adopt here  a simple perturbation expansion in the interaction parameter $\lambda$ appearing in the hamiltonian (\ref{25}). In other words, we look for the coefficients in (\ref{35}) having the form
\be
c_\FF(t)=c_\FF^{(0)}(t)+\lambda c_\FF^{(1)}(t)+\lambda^2 c_\FF^{(2)}(t)+\cdots
\label{38}\en
Each  $c_\FF^{(j)}(t)$ satisfies a differential equation which can be deduced as follows: first we recall that $\Psi(t)$ satisfies the Schr\"odinger equation $i\frac{\partial\Psi(t)}{\partial t}=H(t)\Psi(t)$. Replacing (\ref{35}) in this equation and using the orthonormality of the vectors $\varphi_\FF$'s, we find that
\be
\dot c_{\FF'}(t)=-i\lambda \sum_\FF\,c_\FF(t)\,e^{i(E_{\FF'}-E_\FF)t}\,<\varphi_{\FF'},H_I(t)\varphi_\FF>
\label{39}\en
Replacing now (\ref{38}) in (\ref{39}) we find the following infinite set of differential equations, which we can solve, in principle, up to the desired order in $\lambda$:
\be \left\{\begin{array}{lll}
\dot c_{\FF'}^{(0)}(t)=0,\\
\dot c_{\FF'}^{(1)}(t)=-i\sum_{\FF}\,c_{\FF}^{(0)}(t)\,e^{i(E_{\FF'}-E_\FF)t}\,<\varphi_{\FF'},H_I(t)\varphi_\FF>,\\
\dot c_{\FF'}^{(2)}(t)=-i\sum_{\FF}\,c_{\FF}^{(1)}(t)\,e^{i(E_{\FF'}-E_\FF)t}\,<\varphi_{\FF'},H_I(t)\varphi_\FF>,\\
\ldots\ldots\ldots,\\
\end{array}\right.\label{310}\en
The first equation, together with the assumed initial condition, gives $c_{\FF'}^{(0)}(t)=c_{\FF'}^{(0)}(0)=\delta_{\FF',\FF_0}$. When we replace this solution in the differential equation for $c_{\FF'}^{(1)}(t)$ we get, recalling again that $\Psi(0)=\varphi_{\FF_0}$,
\be
c_{\FF'}^{(1)}(t)=-i\int_0^t\,e^{i(E_{\FF'}-E_{\FF_0})t_1}\,<\varphi_{\FF'},H_I(t_1)\varphi_{\FF_0}>\,dt_1
\label{311}\en
Using this in (\ref{310}) we further get
\be
c_{\FF'}^{(2)}(t)=(-i)^2\,\sum_{\FF}\int_0^t\,\left(\int_0^{t_2}\,e^{i(E_{\FF}-E_{\FF_0})t_1}\,h_{\FF,\FF_0}(t_1)\,
dt_1\right)
e^{i(E_{\FF'}-E_{\FF})t_2}\,h_{\FF',\FF}(t_2)\,dt_2,
\label{312}\en
where we have introduced the shorthand notation
\be
h_{\FF,\GG}(t):=<\varphi_{\FF},H_I(t)\varphi_{\GG}>
\label{313}\en

\subsection{First order corrections}

We continue our analysis computing $P_{\FF_0\rightarrow\FF_f}(t)$ in (\ref{37}) up to the first order corrections in $\lambda$ and assuming that $\FF_f$ is different from $\FF_0$. Hence we have
\be
P_{\FF_0\rightarrow\FF_f}(t)=\left|c_{\FF_f}^{(1)}(t)\right|^2=\lambda^2\left|\int_0^t\,
e^{i(E_{\FF_f}-E_{\FF_0})t_1}\,h_{\FF_f,\FF_0}(t_1)\,dt_1\right|^2
\label{314}\en
Using (\ref{32}) and introducing $\delta E=E_{\FF_f}-E_{\FF_0}$, after some algebra we get
\be
P_{\FF_0\rightarrow\FF_f}(t)=\lambda^2\left(\frac{\delta E\,h/2}{\delta E/2}\right)^2 \left|\sum_{k=0}^{M-1}h_{\FF_f,\FF_0}(t_k)\,e^{it_k\delta E}\right|^2
\label{315}\en
The computation of the matrix elements $h_{\FF_f,\FF_0}(t_k)$ is easily performed. Indeed, because of some standard properties of the bosonic operators, we find that
$$
a_{i,\alpha}^\dagger a_{j,\alpha}\, c_i^{P_{\alpha,k}}\,{c_j^\dagger}{P_{\alpha,k}}\,\varphi_{\FF_0}=
\Gamma_{i,j;\alpha}^{(k)}\varphi_{\FF_{0,k}^{(i,j,\alpha)}}
$$
where
\be
\Gamma_{i,j;\alpha}^{(k)}:=
\sqrt{\frac{(k_j^o+P_{\alpha,k})!}{k_j^o!}\frac{k_i^o!}{(k_i^o-P_{\alpha,k})!}\,n_{j,\alpha}^o\,(1+n_{i,\alpha}^o)}
\label{316}\en
and $\FF_{0,k}^{(i,j,\alpha)}$ differs from $\FF_0$ only for the following replacements: $n_{j,\alpha}^o\rightarrow n_{j,\alpha}^o-1$,
$n_{i,\alpha}^o\rightarrow n_{i,\alpha}^o+1$, $k_{j}^o\rightarrow k_{j}^o+P_{\alpha,k}$, $k_{i}^o\rightarrow k_{i}^o-P_{\alpha,k}$. Notice that in our computations we are implicitly assuming that $k_i^o\geq P_{\alpha,k}$, for all $i$, $k$ and $\alpha$. This is because otherwise the trader $\tau_i$ would have not enough money to buy a share $\sigma_\alpha$!

We find that
\be
h_{\FF_f,\FF_0}(t_k)=2\sum_{i,j,\alpha}p_{i,j}^{(\alpha)}\Gamma_{i,j;\alpha}^{(k)}<\varphi_{\FF_f},
\varphi_{\FF_{0,k}^{(i,j,\alpha)}}>
\label{317}\en
Of course,  due to the orthogonality of the vectors $\varphi_{\FF}$'s, the scalar product $<\varphi_{\FF_f},
\varphi_{\FF_{0,k}^{(i,j,\alpha)}}>$ is different from zero (and equal to one) if and only if $n_{j,\alpha}^f=n_{j,\alpha}^o-1$, $n_{i,\alpha}^f=n_{i,\alpha}^o+1$, $k_{i}^f=k_{i}^o-P_{\alpha,k}$ and $k_{j}^f=k_{j}^o+P_{\alpha,k}$, and all the other {\em new} and {\em old} quantum numbers coincide.

For concreteness sake we now consider two simple situations: in the first example below we just assume that the prices of the various shares do not change with $t$. In the second example we consider the case in which only few changes occur, and we take $M=3$.

\vspace{2mm}

{\bf Example 1: constant prices}

Let us assume that, for all $k$ and for all $\alpha$, $P_{\alpha,k}=P_\alpha(t_k)=P_\alpha$. This means that $\Gamma_{i,j;\alpha}^{(k)}$, $\FF_{0,k}^{(i,j,\alpha)}$ and the related vectors $\varphi_{\FF_{0,k}^{(i,j,\alpha)}}$ do not depend on $k$. Hence $h_{\FF_f,\FF_0}(t_k)$ is also independent of $k$. After few computation we get
\be
P_{\FF_0\rightarrow\FF_f}(t)=\lambda^2\left(\frac{\sin(\delta E t/2)}{\delta E/2}\right)^2 \left|h_{\FF_f,\FF_0}(0)\right|^2
\label{318}\en
to which corresponds a transition probability per unit of time
\be
p_{\FF_0\rightarrow\FF_f}=\lim_{t,\infty}\frac{1}{t}\,P_{\FF_0\rightarrow\FF_f}(t)=2\pi\,\lambda^2\,\delta(E_{\FF_f}-E_{\FF_0}) \left|h_{\FF_f,\FF_0}(0)\right|^2,
\label{319}\en
which shows that, in this limit, a transition between two states is possible only if the two states have the same {free energy}.  The presence of $h_{\FF_f,\FF_0}(0)$
in the final result shows, using our previous remark, that at the order we are considering here a transition is possible only if $\varphi_{\FF_0}$ does not differ from $\varphi_{\FF_f}$ for more than one share in two of the $n_{j,\alpha}$'s and for more than $P_\alpha$ in two of the $k_j$'s. All the other transitions are forbidden.

\vspace{2mm}

{\bf Example 2: few changes in the price}

Let us now fix $M=3$. Formula (\ref{315}) can be rewritten as
\be
P_{\FF_0\rightarrow\FF_f}(t)=4\lambda^2\left(\frac{\sin(\delta E h/2)}{\delta E/2}\right)^2
|\sum_{i,j,\alpha}p_{i,j}^{(\alpha)}(
\Gamma_{i,j;\alpha}^{(0)}<\varphi_{\FF_f},\varphi_{\FF_{0,0}^{(i,j,\alpha)}}>+$$
$$+
\Gamma_{i,j;\alpha}^{(1)}<\varphi_{\FF_f},\varphi_{\FF_{0,1}^{(i,j,\alpha)}}>\,e^{ih\delta E}+
\Gamma_{i,j;\alpha}^{(2)}<\varphi_{\FF_f},\varphi_{\FF_{0,2}^{(i,j,\alpha)}}>\,e^{2ih\delta E})|^2
\label{320}\en
The meaning of this formula  is not very different from the one discussed in the previous example: if we restrict to this order of approximation, the only possibilities for a transition $\FF_0\rightarrow\FF_f$ to occur are those already discussed in Example 1 above. We will see that, in order to get something different, we need to go to higher orders in $\lambda$. In other words, even if the prices depend on time, not new relevant features appear in the transition probabilities.

\vspace{3mm}

As for the validity of the approximation, let us consider the easiest situation: we have constant prices (Example 1) and, moreover, in the summation in (\ref{317}) only one contribution survives, the one with $i_0,j_0$ and $\alpha_0$. Then we have
$h_{\FF_f,\FF_0}(t_0)=2p_{i_0,j_0}^{(\alpha_0)}\Gamma_{i_0,j_0;\alpha_0}<\varphi_{\FF_f},
\varphi_{\FF_{0}^{(i_0,j_0,\alpha_0)}}>$. Because of (\ref{318}), and since our approximation becomes meaningless if  $P_{\FF_0\rightarrow\FF_f}(t)$ exceeds one, it is necessary to have small $\lambda$, small $p_{i,j}^{(\alpha)}$ and large $\delta E$ (if this is possible). However, due to the analytic expression for $\Gamma_{i_0,j_0;\alpha_0}$, see (\ref{316}), we must pay attention to the values of the $n_{j,\alpha}$'s and of the $k_j$'s, since, if they are large, the approximation may likely break down very soon in $t$.

\vspace{2mm}

Let us now finally see what can be said about the portfolio of the trader $\tau_l$. Since we know the initial state of the system, then we know the value of its portfolio at time zero: extending our original definition in (\ref{27}) we have $\hat
\Pi_l(0)=\sum_{\alpha=1}^L P_\alpha(0)\,\hat n_{l,\alpha}(0)+\hat k_l(0)$. As a matter of fact, the knowledge of $\varphi_{\FF_0}$ implies that we know the time zero value of the portfolios of all the traders, clearly! Formula (\ref{315}) gives the transition probability from $\varphi_{\FF_0}$ to $\varphi_{\FF_f}$. This probability is just a single contribution in the computation of the transition probability from a given $\hat \Pi_l(0)$ to a certain $\hat
\Pi_l(t)$, since the same value of the portfolio can be recovered at time $t$ for very many different states $\varphi_{\FF_f}$: all the sets $\GG$ with the same $n_{l,\alpha}^f$ and $k_l^f$ give rise to the same portfolio for $\tau_l$. Hence, if we call $\tilde\FF$ the set of all these sets, we just have to sum up over all these different contributions:
$$
P_{\hat\Pi_l^o\rightarrow\hat\Pi_l^f}(t)=\sum_{\GG\in\tilde\FF}P_{\FF_0\rightarrow\GG}(t)
$$
\subsection{Second order corrections}

We start considering the easiest situation, i.e. the case of a time independent perturbation $H_I$: the prices are constant in time. Hence the integrals in formula (\ref{312}) can be easily computed and the result is the following:
\be
c_{\FF_f}^{(2)}(t)=\sum_{\FF}h_{\FF_f,\FF}(0)h_{\FF,\FF_0}(0){\cal E}_{\FF,\FF_0,\FF_f}(t),
\label{321}\en
where
$$
{\cal E}_{\FF,\FF_0,\FF_f}(t)=\frac{1}{E_\FF-E_{\FF_0}}\left(\frac{e^{i(E_{\FF_f}-E_{\FF_0})t}-1}{E_{\FF_f}-E_{\FF_0}}-
\frac{e^{i(E_{\FF_f}-E_{\FF})t}-1}{E_{\FF_f}-E_{\FF}}\right)
$$
Recalling definition (\ref{313}), we rewrite equation (\ref{321}) as $c_{\FF_f}^{(2)}(t)=\sum_{\FF}<\varphi_{\FF_f},H_I\varphi_{\FF}><\varphi_{\FF},H_I\varphi_{\FF_0}>{\cal E}_{\FF,\FF_0,\FF_f}(t)$ which explicitly shows that up to this order in our perturbation expansion transitions between states which differ, e.g., for 2 shares are allowed: it is enough that some intermediate  state $\varphi_{\FF}$ differs for (plus) one share from $\varphi_{\FF_0}$ and for (minus) one share from $\varphi_{\FF_f}$.

If the $P_\alpha(t)$'s depend on time the situation is a bit more complicated but not very different. Going back to Example 2 above, and considering then a simple (but not trivial) situation in which the prices of the shares really change, we can perform the computation and we find
$$
c_{\FF_f}^{(2)}(t)=(-i)^2\sum_{\FF}\{h_{\FF_f,\FF}(t_0)h_{\FF,\FF_0}(t_0)J_0(\FF,\FF_0,\FF_f;t_1)+$$
$$+
h_{\FF_f,\FF}(t_1)(h_{\FF,\FF_0}(t_0)I_0(\FF,\FF_0;t_1)I_1(\FF_f,\FF;t_2)+h_{\FF,\FF_0}(t_1)J_1(\FF,\FF_0,\FF_f;t_2))+$$
$$+h_{\FF_f,\FF}(t_2)[h_{\FF,\FF_0}(t_0)I_0(\FF,\FF_0;t_1)+h_{\FF,\FF_0}(t_1)I_1(\FF,\FF_0;t_2))I_2(\FF_f,\FF;t_3)+$$
\be+h_{\FF,\FF_0}(t_2)J_2(\FF,\FF_0,\FF_f;t_3)]\},
\label{322}\en
where we have introduced the  functions
$$
I_j(\FF,\GG;t):=\int_{t_j}^t\,e^{i(E_\FF-E_\GG)t'}\,dt'=\frac{1}{i(E_\FF-E_\GG)}
\left(e^{i(E_\FF-E_\GG)t}-e^{i(E_\FF-E_\GG)t_j}\right)
$$
and
$$
J_j(\FF,\GG,\LL;t):=\int_{t_j}^t\,I_j(\FF,\GG;t')\,e^{i(E_\LL-E_\FF)t'}\,dt'.
$$
Needless to say, this last integral could be explicitly computed but we will not show here the explicit result, since it will not be used.

The same comments as above about the possibility of having a non zero transition probability can be repeated also for equation (\ref{322}): it is enough that the time-depending perturbation {\em connect} $\varphi_{\FF_0}$ to $\varphi_{\FF_f}$ via some intermediate state $\varphi_{\FF}$ in a single time sub-interval in order to permit a transition. If this never happens in $[0,t]$, then the transition probability is zero. We can see the problem from a different point of view: if some transition takes place in the interval $[0,t[$, there \underline{must} be another state, $\varphi_{\FF'_f}$, different from $\varphi_{\FF_f}$, such that the transition probability $P_{\FF_0\rightarrow\FF'_f}(t)$ is non zero.

\subsection{Feynman graphs}

Following \cite{mess} we now try to connect the analytic expression of a given approximation of $c_{\FF_f}(t)$ with some kind of Feynman graph in such a way that the higher orders could be easily written  considering a certain set of rules which we will obviously call {\em Feynman rules}.

The starting point is given by the expressions (\ref{311}) and (\ref{312}) for $c_{\FF_f}^{(1)}(t)$ and $c_{\FF_f}^{(2)}(t)$, which is convenient to rewrite in the following form:
\be
c_{\FF_f}^{(1)}(t)=-i\int_0^t\,e^{iE_{\FF_f}t_1}\,<\varphi_{\FF_f},H_I(t_1)\varphi_{\FF_0}>\,e^{-iE_{\FF_0}t_1}\,dt_1
\label{323}\en
and
$$
c_{\FF_f}^{(2)}(t)=(-i)^2\,\sum_{\FF}\int_0^t\,dt_2\int_0^{t_2}\,dt_1\,e^{iE_{\FF_f}t_2}
<\varphi_{\FF_f},H_I(t_2)\varphi_{\FF}>\,e^{-iE_{\FF}t_2}\,\times$$
\be\times e^{iE_{\FF}t_1}\,
<\varphi_{\FF},H_I(t_1)\varphi_{\FF_0}>\,
e^{-iE_{\FF_0}t_1}
\label{324}\en
A graphical way to describe $c_{\FF_f}^{(1)}(t)$ is given in the figure below: at $t=t_0$ the state of the system is $\varphi_{\FF_0}$, which evolves freely (and therefore $e^{-iE_{\FF_0}t_1}\varphi_{\FF_0}$ appears) until the interaction occurs, at $t=t_1$. After the interaction the system is moved to the state $\varphi_{\FF_f}$, which evolves freely (and therefore $e^{-iE_{\FF_f}t_1}\varphi_{\FF_f}$ appears, and the different sign in (\ref{323}) is due to the anti-linearity of the scalar product in the first variable.). The free evolutions are the upward arrows, while the interaction between the initial and the final states, $<\varphi_{\FF_f},H_I(t_1)\varphi_{\FF_0}>$, is described by an horizontal wavy line. Obviously, since the interaction may occur at any time between 0 and $t$, we have to integrate on all these possible $t_1$'s and multiply the result for $-i$.

\vspace{2cm}

\begin{picture}(220,170)(-130,0)

\put(-40,45){\thicklines\vector(0,1){170}}

\put(-50,50){\makebox(0,0){$t_0$}}
\put(-50,130){\makebox(0,0){$t_1$}}
\put(-50,200){\makebox(0,0){$t$}}

\put(130,50){\vector(-1,2){30}}
\put(105,100){\line(-1,2){10}}

\put(95,120){\vector(1,2){10}}
\put(105,140){\line(1,2){30}}

\put(95,120){\line(-1,0){5}}
\put(5,120){\line(1,0){5}}

\qbezier[100](10,120)(11,123)(14,124)
\qbezier[100](14,124)(17,123)(18,120)
\qbezier[100](18,120)(19,117)(22,116)
\qbezier[100](22,116)(25,117)(26,120)

\qbezier[100](26,120)(27,123)(30,124)
\qbezier[100](30,124)(31,123)(34,120)
\qbezier[100](34,120)(35,117)(38,116)
\qbezier[100](38,116)(41,117)(42,120)

\qbezier[100](42,120)(43,123)(46,124)
\qbezier[100](46,124)(49,123)(50,120)
\qbezier[100](50,120)(51,117)(54,116)
\qbezier[100](54,116)(57,117)(58,120)

\qbezier[100](58,120)(59,123)(62,124)
\qbezier[100](62,124)(65,123)(66,120)
\qbezier[100](66,120)(67,117)(70,116)
\qbezier[100](70,116)(73,117)(74,120)

\qbezier[100](74,120)(75,123)(78,124)
\qbezier[100](78,124)(81,123)(82,120)
\qbezier[100](82,120)(83,117)(86,116)
\qbezier[100](86,116)(89,117)(90,120)

\put(150,50){\makebox(0,0){$\varphi_{\FF_0}$}}
\put(150,200){\makebox(0,0){$\varphi_{\FF_f}$}}
\put(160,120){\makebox(0,0){$<\varphi_{\FF_f},H_I(t_1)\varphi_{\FF_0}>$}}

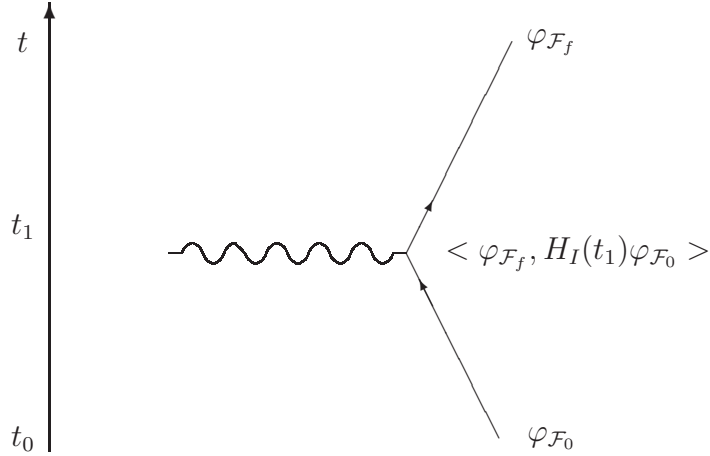
\put(100,0){\makebox(0,0){Figure 1: graphical expression for $c_{\FF_f}^{(1)}(t)$}}

\end{picture}

\vspace{1cm}

In a similar way we can construct the Feynman graph for $c_{\FF_f}^{(2)}(t)$, $c_{\FF_f}^{(3)}(t)$ and so on. For example $c_{\FF_f}^{(2)}(t)$ can be deduced by a graph like the one in Figure 2, where two interactions occur, the first at $t=t_1$ and the second at $t=t_2$:

\vspace{2cm}

\begin{picture}(220,170)(-130,0)

\put(-40,45){\thicklines\vector(0,1){190}}

\put(-50,50){\makebox(0,0){$t_0$}}
\put(-50,120){\makebox(0,0){$t_1$}}
\put(-50,170){\makebox(0,0){$t_2$}}
\put(-50,220){\makebox(0,0){$t$}}

\put(130,50){\vector(-1,2){30}}
\put(105,100){\line(-1,2){10}}

\put(95,120){\vector(1,2){10}}
\put(105,140){\line(1,2){10}}

\put(115,160){\vector(-1,2){10}}
\put(106,178){\line(-1,2){20}}


\put(95,120){\line(-1,0){5}}
\put(5,120){\line(1,0){5}}

\qbezier[100](10,120)(11,123)(14,124)
\qbezier[100](14,124)(17,123)(18,120)
\qbezier[100](18,120)(19,117)(22,116)
\qbezier[100](22,116)(25,117)(26,120)

\qbezier[100](26,120)(27,123)(30,124)
\qbezier[100](30,124)(31,123)(34,120)
\qbezier[100](34,120)(35,117)(38,116)
\qbezier[100](38,116)(41,117)(42,120)

\qbezier[100](42,120)(43,123)(46,124)
\qbezier[100](46,124)(49,123)(50,120)
\qbezier[100](50,120)(51,117)(54,116)
\qbezier[100](54,116)(57,117)(58,120)

\qbezier[100](58,120)(59,123)(62,124)
\qbezier[100](62,124)(65,123)(66,120)
\qbezier[100](66,120)(67,117)(70,116)
\qbezier[100](70,116)(73,117)(74,120)

\qbezier[100](74,120)(75,123)(78,124)
\qbezier[100](78,124)(81,123)(82,120)
\qbezier[100](82,120)(83,117)(86,116)
\qbezier[100](86,116)(89,117)(90,120)


\put(115,160){\line(1,0){5}}
\put(200,160){\line(1,0){5}}

\qbezier[100](120,160)(121,163)(124,164)
\qbezier[100](124,164)(127,163)(128,160)
\qbezier[100](128,160)(129,157)(132,156)
\qbezier[100](132,156)(135,157)(136,160)

\qbezier[100](136,160)(137,163)(140,164)
\qbezier[100](140,164)(141,163)(144,160)
\qbezier[100](144,160)(145,157)(148,156)
\qbezier[100](148,156)(151,157)(152,160)

\qbezier[100](152,160)(153,163)(156,164)
\qbezier[100](156,164)(159,163)(160,160)
\qbezier[100](160,160)(161,157)(164,156)
\qbezier[100](164,156)(167,157)(168,160)

\qbezier[100](168,160)(169,163)(172,164)
\qbezier[100](172,164)(175,163)(176,160)
\qbezier[100](176,160)(177,157)(180,156)
\qbezier[100](180,156)(183,157)(184,160)

\qbezier[100](184,160)(185,163)(188,164)
\qbezier[100](188,164)(191,163)(192,160)
\qbezier[100](192,160)(193,157)(196,156)
\qbezier[100](196,156)(199,157)(200,160)

\put(150,50){\makebox(0,0){$\varphi_{\FF_0}$}}
\put(110,220){\makebox(0,0){$\varphi_{\FF_f}$}}
\put(95,140){\makebox(0,0){$\varphi_{\FF}$}}
\put(160,120){\makebox(0,0){$<\varphi_{\FF},H_I(t_1)\varphi_{\FF_0}>$}}
\put(55,160){\makebox(0,0){$<\varphi_{\FF_f},H_I(t_2)\varphi_{\FF}>$}}

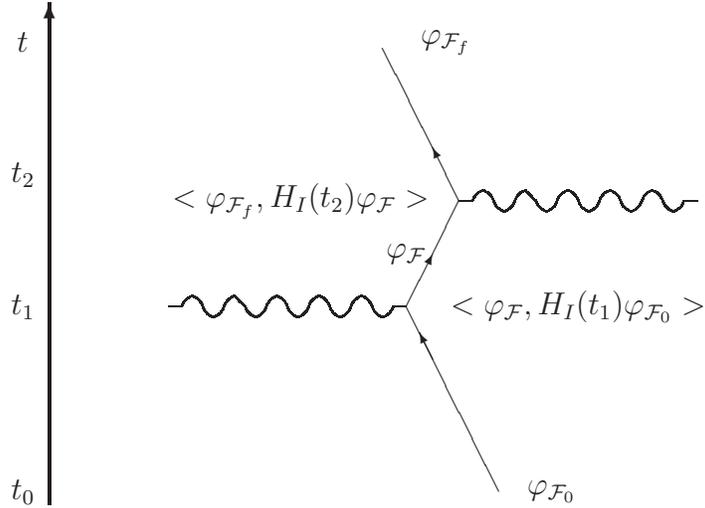
\put(100,0){\makebox(0,0){Figure 2: graphical expression for $c_{\FF_f}^{(2)}(t)$}}

\end{picture}

\vspace{1cm}
Because of the double interaction we have to integrate twice the result, since $t_1\in(0,t_2)$ and $t_2\in(0,t)$. For the same reason we have to sum  over all the possible intermediate states, $\varphi_{\FF}$. The free time evolution for the various free fields also appear, as well as a $(-i)^2$. Following these same rules we could also give at least a formal expression for the other coefficients, as $c_{\FF_f}^{(3)}(t)$, $c_{\FF_f}^{(4)}(t)$ and so on: the third order correction $c_{\FF_f}^{(3)}(t)$ contains, for instance, a double sum on the intermediate states, allowing in this way a transition from a state with, say, $n_{i,\alpha}^o$ shares to a state with $n_{i,\alpha}^f=n_{i,\alpha}^o+3$ shares, a triple time integral and a factor $(-i)^3$.

\section{Conclusions}
We have shown how quantum statistical dynamics can be adopted to construct and analyze simplified models of closed stock markets, where no derivatives are considered. In particular we have shown that both the Schr\"odinger and the Heisenberg pictures can be successfully used in the perturbative analysis of the time evolution of the market: however, the approximations considered in the Heisenberg picture are not completely under control because of the many assumptions adopted as it happens using the Schr\"odinger wave function of the market and looking for transition probabilities.

We have also shown that the Feynman graphs technique can be adopted for the perturbative analysis of our market, and some simple rules to write down the integral analytic expression for the transition probabilities have been deduced.

Of course a more detailed analysis of the model should be performed, in particular looking for those adjustments which can make more realistic the market we have described so far. These should include for instance source and sink effects to mimikate non conservation of the number of shares, short terms exchanges, financial derivatives and so on. Incidentally, we believe that the hamiltonian $H$ in (\ref{25}) and the related differential equations in (\ref{28}), could be interesting by themselves, hopefully in  the description of some (realistic?) many-body system. We hope to consider this aspect in the next future.

\section*{Acknowledgements}

This work has been financially supported in part by M.U.R.S.T.,
within the  project {\em Problemi Matematici Non Lineari di
Propagazione e Stabilit\`a nei Modelli del Continuo}, coordinated by
Prof. T. Ruggeri.

\end{document}